\documentclass[aps,prb,reprint,groupedaddress]{revtex4-2}

\usepackage{amsmath}
\usepackage{amssymb}
\usepackage{mathrsfs}

\usepackage{txfonts}
\usepackage{graphicx}

\begin{document}

\title{Frustrations and orderings in Ising chain with multiple interactions}

\author{A.V. Zarubin}
\email{Alexander.Zarubin@imp.uran.ru}
\affiliation{M.N. Mikheev Institute of Metal Physics of Ural Branch of Russian Academy of Sciences, 620108 Ekaterinburg, Russia}

\author{F.A. Kassan-Ogly}
\email{Felix.Kassan-Ogly@imp.uran.ru}
\affiliation{M.N. Mikheev Institute of Metal Physics of Ural Branch of Russian Academy of Sciences, 620108 Ekaterinburg, Russia}

\author{A.I. Proshkin}
\affiliation{M.N. Mikheev Institute of Metal Physics of Ural Branch of Russian Academy of Sciences, 620108 Ekaterinburg, Russia}


\begin{abstract}
The frustration properties of the Ising model on a one-dimensional monoatomic equidistant lattice are investigated taking into account the exchange interactions of atomic spins at the sites of the first (nearest), second (next-nearest) and third neighbors. The exact solution of the model was obtained using the Kramers--Wannier transfer matrix method. The types of magnetic ordering of the ground state of the model are determined, and a magnetic phase diagram is constructed. Criteria are formulated for the occurrence of magnetic frustrations in the presence of competition between the energies of exchange interactions. Non-zero entropy values in the ground state of the frustrated system were found.
\end{abstract}

\maketitle

\section{Introduction}
\label{intro}

Spin systems with magnetic frustrations are promising objects, in which incommensurate, helicoidal, chiral, and other exotic magnetic structures are observed, as well as special magnetic states, such as spin liquid and spin ice~\cite{Kassan-Ogly:2010:,Diep:2013,Vasiliev:2018:}. With a sufficiently large experimental data, a number of properties of frustrated systems do not yet have a theoretical description.

The frustration properties of the Ising model on a one-dimensional monoatomic equidistant lattice are studied taking into account the exchange interactions of atomic spins at the sites of the first (nearest), second (next-nearest), and third neighbors.

This model may have an exact solution, which makes it possible to qualitatively consider the desired characteristics, including explaining the properties of magnetic materials caused by frustrations not available for description in the framework of the perturbation theory~\cite{Baxter:1982}.

Note that the Ising model has long been widely used in the theory of magnetism and has a set of well-known solutions, but it has no a systematic description of its frustration properties.

\section{Thermodynamic functions of the Ising chain}
\label{sec:TD}

The one-dimensional Ising model taking into account the exchange interactions of atomic spins at sites of the first (nearest), second (next-nearest) and third neighbors is defined by a Hamiltonian of form
\begin{equation}
\mathscr{H}=-\sum_{p=1}^{b}\sum_{n=1}^{N-p}J_{p}\sigma_{n}\sigma_{n+p},
\label{eq:H:0}
\end{equation}
where $b$ is the number of exchange interactions of the chain spins in the problem (in this case $b=3$), $J_{1}$ is the parameter of the exchange interaction between the spins at the nearest sites of the linear lattice, $J_{2}$ is the parameter of the exchange interaction between the spins at the next-nearest lattice sites, $J_{3}$ is the exchange interaction parameter between the spins on the third lattice sites, the symbol $\sigma_{n}$ denotes the $z$-projection of the spin operator $\sigma=\pm1$ of the atom at site $n$, and $N$ is the number of chain sites spins.

It should be noted that in further transformations, we will set quantities such as the Boltzmann constant ($k_{\text{B}}$) equal to unity, while quantities $T$, $J_{2}$, and $J_{3}$ will be measured in the units of $|J_{1}|$ as is usually done in the theory of low-dimensional systems.

In the method of the Kramers--Wannier transfer matrix \cite{Kramers:1941:1} and with the imposition of the Born--von Karman cyclic boundary conditions, the partition function is
\begin{equation}
Z=\lambda_{1}^{N},
\label{eq:PF}
\end{equation}
where $\lambda_{1}$ is the principal (the only one maximal real) eigenvalue of the transfer matrix, which for this type of matrix always exists according to the Frobenius--Perron theorem~\cite{Domb:1960}.

In this case, for Hamiltonian (\ref{eq:H:0}) the transfer matrix can be reduced to a block form and get a characteristic equation that defines the principal eigenvalue,
\begin{equation}
\lambda^{4}+a_{3}\lambda^{3}+a_{2}\lambda^{2}+a_{1}\lambda+a_{0}=0,
\label{eq:N3:CP1}
\end{equation}
where the coefficients are 
\[
a_{3}=-2e^{K_{2}}\cosh(K_{1}+K_{3}),\quad a_{2}=2\sinh(2K_{2}),
\]
\[
a_{1}=4e^{-K_{2}}\sinh(2K_{3})\sinh(K_{1}-K_{3}),\quad a_{0}=4\sinh^{2}(2K_{3}),
\]
which are represented by dimensionless variables
\[
K_{1,2,3}=\beta J_{1,2,3},\quad\beta=\frac{1}{k_{\text{B}}T}.
\]
For the fourth order equation, the principal eigenvalue of the transfer matrix is expressed in radicals and has the following form
\begin{equation}
\lambda_{1}=-\frac{a_{3}}{4}-\Psi+\frac{1}{2}\sqrt{-4\Psi^{2}-2p+\frac{q}{S}},
\label{eq:N3:L1}
\end{equation}
\[
p=a_{2}-\frac{3}{8}a_{3}^{2},\quad q=a_{1}-\frac{a_{2}a_{3}}{2}+\frac{a_{3}^{3}}{8},
\]
\[
\Psi=\frac{1}{2}\sqrt{-\frac{2}{3}p+\frac{1}{3}\left(\Theta+\frac{\Delta_{0}}{\Theta}\right)},
\quad
\Theta=\sqrt[3]{\frac{\Delta_{1}+\sqrt{\Delta_{1}^{2}-4\Delta_{0}^{3}}}{2}},
\]
\[
\Delta_{0}=12a_{0}-3a_{1}a_{3}+a_{2}^{2},
\]
\[
\Delta_{1}=-72a_{0}a_{2}+27a_{0}a_{3}^{2}+27a_{1}^{2}-9a_{1}a_{2}a_{3}+2a_{2}^{3}.
\]
In the case of taking into account the exchange interaction of spins at the sites of only the first and second neighbors ($J_{3}=0$), the principal eigenvalue (\ref{eq:N3:L1}) corresponds to
\begin{equation}
\lambda_{1}=e^{K_{2}}\left(\cosh K_{1}+\sqrt{\sinh^{2}K_{1}+e^{-4K_{2}}}\right),
\label{eq:N2:L1}
\end{equation}
and when taking into account the exchange interaction of spins on the sites of only the first neighbors ($J_{2}=J_{3}=0$), the principal eigenvalue
\begin{equation}
\lambda_{1}=2\cosh K_{1}.
\label{eq:N1:L1}
\end{equation}

As a result, all thermodynamic functions of the system, including the Helmholtz free energy, internal energy, entropy, heat capacity per spin, are determined only through the principal eigenvalue of the transfer matrix~\cite{Baxter:1982}.

\section{Magnetic phase diagram of the system ground state}
\label{sec:PD}

The magnetic phase diagram of the ground state of such a model is determined by the behavior of the minimum energy of the spin system configurations at zero temperature, depending on the model parameters. Configuration energy is defined as the internal energy per spin
\[
U=\frac{T^{2}}{\lambda_{1}}\frac{\partial\lambda_{1}}{\partial T}
\]
at zero temperature.

\begin{figure}[hbt]
\noindent\centering
\includegraphics[width=0.85\columnwidth,clip]{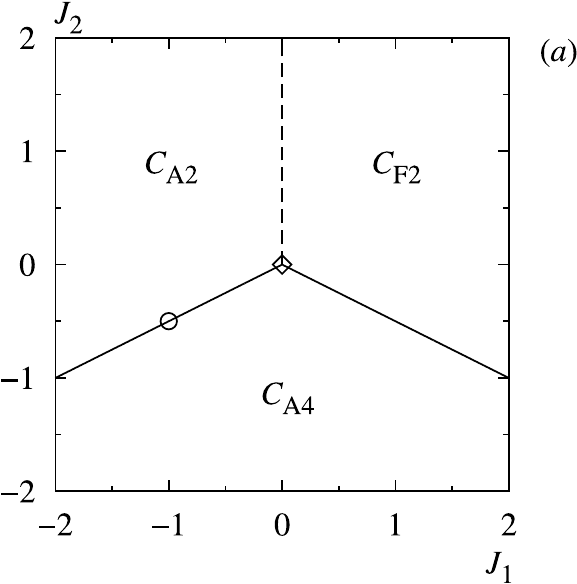}\quad
\includegraphics[width=0.85\columnwidth,clip]{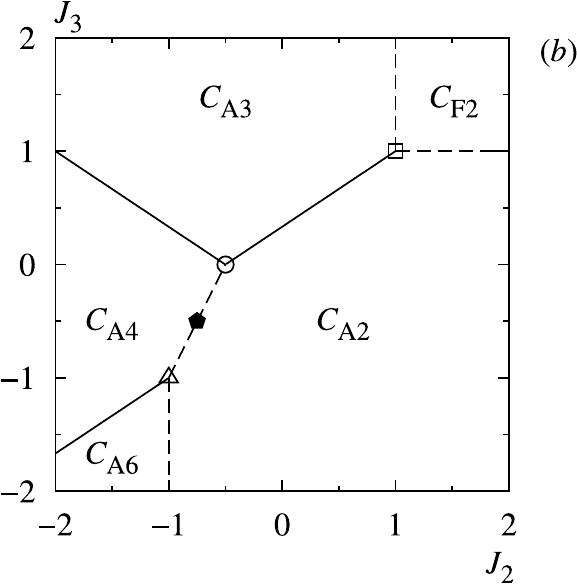}
\caption{Magnetic phase diagram of the ground state of the one-dimensional Ising model taking into account exchange interactions of spins in the lattice sites of the first (nearest), second (next-nearest) neighbors (a), and third neighbors (b) with the antiferromagnetic interaction of the nearest neighbors ($J_{1}=-1$)}
\label{fig:PD}
\end{figure}

In the case of taking into account the spin interactions of only the first and second neighbors in the ground state, three configurations are formed~\cite{Katsura:1972,Pokrovskii:1982:,Price:1984}, which are shown in the magnetic phase diagram of the ground state (see figure~\ref{fig:PD}a).

The first type of spin configurations is characterized by antiferromagnetic ordering and is designated $C_{\text{A}2}$. The second type of spin configurations is characterized by ferromagnetic ordering with the designation $C_{\text{F}2}$. The third type is determined by magnetic ordering with a quadrupling of the chain translation period with the designation $C_{\text{A}4}$.

By determining the energies of these configurations, one can obtain the ratios of the exchange interaction parameters of the model, at which the structure of the ordering of the spin configurations of the ground state is rearranged,
\[
J_{1}=\begin{cases}
2J_{2}, & J_{1}<0\land J_{2}\leqslant0,\\
-2J_{2},\quad & J_{1}>0\land J_{2}\leqslant0,\\
0, & J_{2}\geqslant0.
\end{cases}
\]
In this case, the magnetic phase diagram of the ground state (see figure~\ref{fig:PD}a) contains four variants of the relations of the parameters of exchange interactions between the spins at the sites of the first and second neighbors of the Ising chain, and only two cases
\begin{equation}
(J_{1}>0,J_{2}<0),\quad(J_{1}<0,J_{2}<0).
\label{eq:N2:J1:J2:m}
\end{equation}
describe a system with competing exchange interactions. It is in these two areas of the magnetic phase diagram that frustrations exist~\cite{Zarubin:2019:}.

In the case of taking into account the interaction of spins at the sites of the first, second, and third neighbors, depending on the signs of the parameters of the exchange interactions of the spins of the chain in the ground state, five types of possible spin configurations with the minimum energy are realized (see figure~\ref{fig:PD}b).

In addition to the above-mentioned configurations, there are also configurations characterized by magnetic ordering with tripling ($C_{\text{A}3}$) and with sixtupling ($C_{\text{A}6}$) of the chain translation period.

The relations of the exchange interaction parameters of the model at which the structure of the ordering of the spin configurations of the ground state is rearranged are
\[
J_{1}=\begin{cases}
-J_{3}, & J_{2}>J_{3}\land J_{2}>-J_{3},\\
2J_{2}-3J_{3}, & J_{2}\leqslant J_{3}\land J_{3}>0,\\
2J_{2}-J_{3}, & J_{2}<J_{3}\land J_{3}<0,\\
J_{2}, & J_{2}>J_{3}\land J_{2}<-J_{3},\\
-J_{2}, & J_{2}<J_{3}\land J_{2}>-J_{3},\\
2J_{2}+3J_{3}, & J_{2}\leqslant-J_{3}\land J_{3}>0,\\
-2J_{2}+3J_{3},\quad & J_{2}\leqslant J_{3}\land J_{3}<0,\\
-2J_{2}-J_{3}, & J_{2}<-J_{3}\land J_{3}>0,\\
-2J_{2}-3J_{3}, & J_{2}\leqslant-J_{3}\land J_{3}<0,
\end{cases}
\]
with the formation of a complicated structure of the boundaries of the regions of these configurations, as shown in the magnetic phase diagram of the spin system \cite{Price:1983,Barreto:1985,Selke:1985} in figure~\ref{fig:PD}b.

Note that the magnetic phase diagram shown in figure~\ref{fig:PD}b, is symmetric with respect to the replacement of signs of exchange interaction parameters
\begin{equation}
\{J_{1},J_{3}\}\Leftrightarrow\{-J_{1},-J_{3}\}.
\label{eq:J1:J3}
\end{equation}
Therefore, using this thermodynamic symmetry, for further analysis of the model, it is sufficient to consider its behavior only with one sign of the exchange interaction parameter between the spins at the sites of the nearest neighbors.

The considered model contains eight variants of the relations of the exchange interaction parameters of first, second, and third neighbors of the chain spins. The presented variants of the parameters
\begin{equation}
(J_{1}<0,J_{2}>0,J_{3}>0),\quad(J_{1}>0,J_{2}>0,J_{3}<0),
\label{eq:N3:Q:14}
\end{equation}
\begin{equation}
(J_{1}<0,J_{2}<0,J_{3}<0),\quad(J_{1}>0,J_{2}<0,J_{3}>0),
\label{eq:N3:Q:32}
\end{equation}
\begin{equation}
(J_{1}<0,J_{2}<0,J_{3}>0),\quad(J_{1}>0,J_{2}<0,J_{3}<0)
\label{eq:N3:Q:23}
\end{equation}
define a system with competing exchange interactions between spins and in frustration regimes.

\section{Entropy of the system in the frustration modes}
\label{sec:EN}

On the magnetic phase diagram of the ground state of the Ising chain with competing spin exchange interactions (figure~\ref{fig:PD}) in regions outside of the boundaries of the spin configurations and at the boundaries marked by dashed lines, the corresponding entropy values
\begin{equation}
S=\ln\lambda_{1}+\frac{T}{\lambda_{1}}\frac{\partial\lambda_{1}}{\partial T}
\label{eq:S0}
\end{equation}
at zero temperature are zero
\begin{equation}
\lim_{T\to0}S=0.
\label{eq:S0:0}
\end{equation}
At these boundaries, a rearrangement of the ordering of the ground state occurs, and a number of configurations of the system with the minimum energy is equal to the sum of the configurations of the regions adjacent to the boundary.

At the boundaries indicated by solid lines (figure~\ref{fig:PD}), the entropy of the ground state is nonzero,
\begin{equation}
\lim_{T\to0}S>0,
\label{eq:S0:fr}
\end{equation}
and the number of configurations of the system with the minimum energy is greater than the sum of the configurations of the adjacent regions of the phase diagram. Such a set of spin configurations of the system at zero temperature is associated with the rearrangement of the magnetic structure and the occurrence of an infinite set of spin configurations in the thermodynamic limit at this phase point, including those with the violation of translational invariance. In the terminology of \cite{Pokrovskii:1982:,Selke:1985}, such lines are called multiphase lines. Also found on the phase diagram, only triple points (mutiphase point \cite{Fisher:1981}) were detected in accordance with the Gibbs phase rule.

Such states of the system in which the entropy of the ground state is non-zero (\ref{eq:S0:fr}) should be referred to as \emph{frustrated}. Also note that this result by no means contradicts the third law of thermodynamics~\cite{Zarubin:2019:}.

If there is no exchange interactions between the spins ($J_{i}=0$), at the point on the phase diagram marked by diamond in figure~\ref{fig:PD}a, a paramagnetic state is realized, characterized by the fact that all system configurations have the same probability and have the same energy. The entropy of such a state of the system is equal to the natural logarithm of the two, 
\begin{equation}
S=\ln2\approx0.693,
\label{eq:N3:S0:0}
\end{equation}
and is the same (maximal) at any temperature. Since for any values of the exchange interaction parameters, the entropy of an infinitely large temperature is also equal to 
\begin{equation}
\lim_{T\to\infty}S=\ln2,
\label{eq:S0:T8}
\end{equation}
hence it is clear that the Ising paramagnet is an absolutely frustrated system~\cite{Zarubin:2019:}.

\begin{figure}[ht]
\centering\includegraphics[width=0.85\columnwidth,clip]{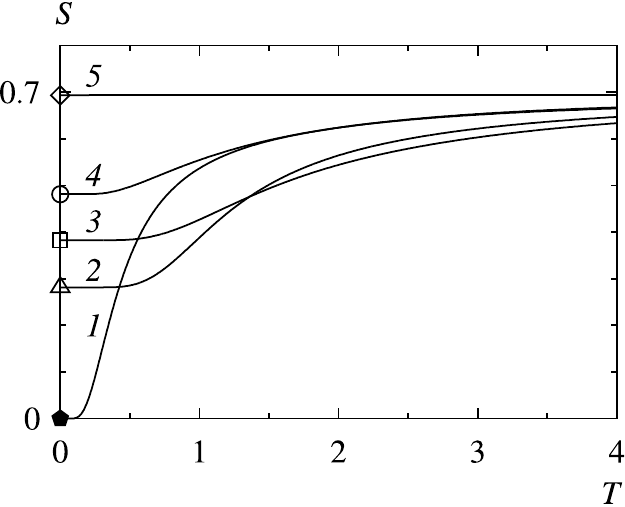}
\caption{The entropy of the Ising chain (\ref{eq:S0}) at the following points in the phase space, where the ratios of the exchange interaction parameters  ($J_{1}=-1$, $J_{2}=-3/4$, $J_{3}=-1/2$) (line~1), ($J_{1}=J_{2}=J_{3}=-1$) (line~2), ($J_{1}=-J_{2}=-J_{3}=-1$) (line~3), ($J_{1}=-1$, $J_{2}=-1/2$, $J_{3}=0$) (line~4), and ($J_{1}=J_{2}=J_{3}=0$) (line~5)}
\label{fig:N3:S0:5} 
\end{figure}

\begin{figure}[ht]
\centering\includegraphics[width=0.85\columnwidth,clip]{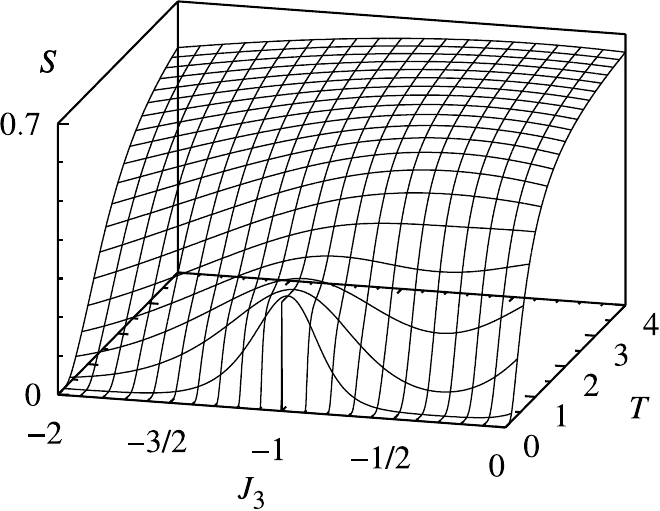}
\caption{Entropy of the Ising chain (\ref{eq:S0}) with the antiferro-antiferro-antiferromagnetic variant of the exchange interaction parameters ($J_{1}=J_{2}=-1$ and $J_{3}<0$)}
\label{fig:S0:mmm} 
\end{figure}

\begin{figure}[ht]
\centering \includegraphics[width=0.85\columnwidth,clip]{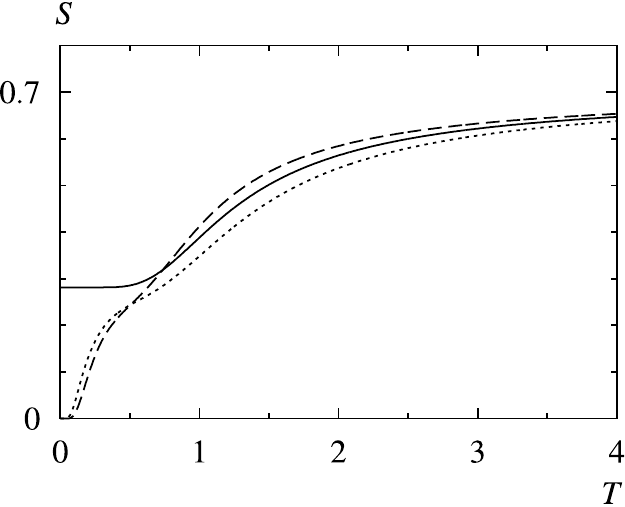}
\caption{The entropy of the Ising chain (\ref{eq:S0}) in the vicinity of the frustration point of the system, where the antiferro-antiferro-antiferromagnetic variant of the exchange interaction parameters are $J_{1}=J_{2}=-1$, and the values of the parameter $J_{3}=\{-6/5,-1,-4/5\}$ correspond to the dotted, solid and dashed lines in the graph}
\label{fig:S0:mmm:} 
\end{figure}

When the ratio of the parameters of the exchange interaction of model $J_{2}=-|J_{1}|/2$ and $J_{3}=0$ at the point marked on the phase diagram (figure~\ref{fig:PD}) by a circle, the entropy at zero temperature is equal to the natural logarithm of the golden ratio,
\begin{equation}
\lim_{T\to0}S=\ln\frac{1+\sqrt{5}}{2}\approx0.481.
\label{eq:N3:S0:1}
\end{equation}

In the phase diagram (figure~\ref{fig:PD}) thick solid lines mark the boundaries of the regions of spin configurations with the ratios of exchange interaction parameters
\[
J_{3}=-\frac{J_{1}-2J_{2}}{3},\quad J_{1}/2<J_{2}\leqslant-J_{1},
\]
and the points are marked by squares ($J_{2}=|J_{1}|$ and $J_{3}=-J_{1}$), in which the temperature zero entropy is 
\begin{multline*}
\lim_{T\to0}S=\ln\left[\frac{1}{3}\left(1+\sqrt[3]{\frac{3^{3}+2+\sqrt{(3^{3}+2^{2})3^{3}}}{2}}\right.\right.\\
+\left.\left.\sqrt[3]{\frac{2}{3^{3}+2+\sqrt{(3^{3}+2^{2})3^{3}}}}\right)\right]\approx0.382,
\label{eq:N3:S0:2}
\end{multline*}
Also, thin solid lines mark the boundaries of the configurations with a ratio of model parameters
\[
J_{3}=\frac{J_{1}-2J_{2}}{3},\quad J_{2}\leqslant-J_{1},\quad J_{2}<J_{1}/2,
\]
\[
J_{3}=\frac{J_{1}+2J_{2}}{3},\quad J_{2}\leqslant J_{1},\quad J_{2}<-J_{1}/2,
\]
and triangular marks indicate the points ($J_{2}=-|J_{1}|$ and $J_{3}=J_{1}$) in the phase diagram (figure~\ref{fig:PD}), in which the zero-temperature entropy has the following meaning
\begin{multline*}
\lim_{T\to0}S=\ln\left[\frac{1}{3}\left(\sqrt[3]{\frac{3^{3}+\sqrt{(3^{3}-2^{2})3^{3}}}{2}}\right.\right.\\
+\left.\left.3\sqrt[3]{\frac{2}{3^{3}+\sqrt{(3^{3}-2^{2})3^{3}}}}\right)\right]\approx0.281,
\label{eq:N3:S0:3}
\end{multline*}
Note that the frustration positions of the system on the phase diagram correspond to multiphase lines.

The presented variety of temperature-related entropy values and their temperature evolution are shown in figure~\ref{fig:N3:S0:5}, and the positions of the entropy of the ground state are indicated by labels corresponding to figure~\ref{fig:PD}.

An example of the entropy temperature behavior depending on the exchange interaction parameter is shown in figure~\ref{fig:S0:mmm}. In the vicinity of the frustration position, the behavior of the entropy is shown in figure~\ref{fig:S0:mmm:}.

\section{Conclusions}

Using the Kramers--Wannier transfer matrix method, exact analytical expressions for the thermodynamic quantities of the Ising chain are obtained.

The behavior of the zero-temperature internal energy is calculated, and the magnetic phase diagram of the ground state is determined taking into account the interaction of spins at the sites of the first (nearest), second (next-nearest), and third neighbors. The existence of five spin configurations in the ground state was found.

The analysis of the configuration features of the ground state and the frustration properties of the system is carried out. In this phase diagram, only triple points were detected in accordance with the Gibbs phase rule.

It is shown that, in the frustration mode, the system experiences a rearrangement of the magnetic ordering structure of the ground state, which includes a large set of spin configurations comparable to the size of the system, including those without any translational invariance.

The characteristic behavior of the entropy of the system in the frustration points and near them is analyzed, a fundamental difference in the behavior of the magnetic system in the frustration region and outside of it is shown.

The criteria are formulated and the relations of the model parameters are determined at which magnetic frustrations arise due to the competition between the energies of the exchange interactions of the spins in the one-dimensional systems under consideration.

It was determined that the most important sign of the existence of magnetic frustrations in the system is the nonzero value of zero-temperature entropy in this mode, and this property by no means contradicts the third law of thermodynamics. The zero-temperature entropy is shown to have the nonzero values both at all triple points and at some phases boundaries.

The obtained exact solutions for all thermodynamic functions of the system make it possible to determine the effect of frustrations on the experimental observables~\cite{Zarubin:2019:,Zarubin:2016}.

Note that the features of the behavior of magnetic systems with frustration are observed in real antiferromagnets based on rare earth metals and actinide compounds, as well as in a number of organometallic polymers, molecular and quasi-one-dimensional frustrated magnets (see the discussion in Ref.~\cite{Zarubin:2019:}).

Thus, the proposed analysis scheme allows us to consider a wide range of phenomena in one-dimensional (or quasi-one-dimensional) magnetic systems with frustrations and to describe their connection with the features of the thermodynamic functions.

\begin{acknowledgments}
The research was carried out within the state assignment of Minobrnauki of Russia (theme Quantum No. AAAA-A18-118020190095-4), supported in part by Ural Branch of the Russian Academy of Sciences (project No. 18-2-2-11).
\end{acknowledgments}

\bibliographystyle{apsrev4-2}
\bibliography{zarubin2_arxiv}

\begin{thebibliography}{15}%
\makeatletter
\providecommand \@ifxundefined [1]{%
 \@ifx{#1\undefined}
}%
\providecommand \@ifnum [1]{%
 \ifnum #1\expandafter \@firstoftwo
 \else \expandafter \@secondoftwo
 \fi
}%
\providecommand \@ifx [1]{%
 \ifx #1\expandafter \@firstoftwo
 \else \expandafter \@secondoftwo
 \fi
}%
\providecommand \natexlab [1]{#1}%
\providecommand \enquote  [1]{``#1''}%
\providecommand \bibnamefont  [1]{#1}%
\providecommand \bibfnamefont [1]{#1}%
\providecommand \citenamefont [1]{#1}%
\providecommand \href@noop [0]{\@secondoftwo}%
\providecommand \href [0]{\begingroup \@sanitize@url \@href}%
\providecommand \@href[1]{\@@startlink{#1}\@@href}%
\providecommand \@@href[1]{\endgroup#1\@@endlink}%
\providecommand \@sanitize@url [0]{\catcode `\\12\catcode `\$12\catcode
  `\&12\catcode `\#12\catcode `\^12\catcode `\_12\catcode `\%12\relax}%
\providecommand \@@startlink[1]{}%
\providecommand \@@endlink[0]{}%
\providecommand \url  [0]{\begingroup\@sanitize@url \@url }%
\providecommand \@url [1]{\endgroup\@href {#1}{\urlprefix }}%
\providecommand \urlprefix  [0]{URL }%
\providecommand \Eprint [0]{\href }%
\providecommand \doibase [0]{https://doi.org/}%
\providecommand \selectlanguage [0]{\@gobble}%
\providecommand \bibinfo  [0]{\@secondoftwo}%
\providecommand \bibfield  [0]{\@secondoftwo}%
\providecommand \translation [1]{[#1]}%
\providecommand \BibitemOpen [0]{}%
\providecommand \bibitemStop [0]{}%
\providecommand \bibitemNoStop [0]{.\EOS\space}%
\providecommand \EOS [0]{\spacefactor3000\relax}%
\providecommand \BibitemShut  [1]{\csname bibitem#1\endcsname}%
\let\auto@bib@innerbib\@empty
\bibitem [{\citenamefont {Kassan-Ogly}\ and\ \citenamefont
  {Filippov}(2010)}]{Kassan-Ogly:2010:}%
  \BibitemOpen
  \bibfield  {author} {\bibinfo {author} {\bibfnamefont {F.~A.}\ \bibnamefont
  {Kassan-Ogly}}\ and\ \bibinfo {author} {\bibfnamefont {B.~N.}\ \bibnamefont
  {Filippov}},\ }\href {https://doi.org/10.3103/S1062873810100394} {\bibfield
  {journal} {\bibinfo  {journal} {Bull. {R}uss. {A}cad. {S}ci. {P}hys.}\
  }\textbf {\bibinfo {volume} {74}},\ \bibinfo {pages} {1452} (\bibinfo {year}
  {2010})}\BibitemShut {NoStop}%
\bibitem [{\citenamefont {Diep}(2013)}]{Diep:2013}%
  \BibitemOpen
  \bibinfo {editor} {\bibfnamefont {H.~T.}\ \bibnamefont {Diep}},\ ed.,\ \href
  {https://doi.org/10.1142/8676} {\emph {\bibinfo {title} {Frustrated spin
  systems}}},\ \bibinfo {edition} {2nd}\ ed.\ (\bibinfo  {publisher} {World
  {S}cientific},\ \bibinfo {address} {New {J}ersey},\ \bibinfo {year}
  {2013})\BibitemShut {NoStop}%
\bibitem [{\citenamefont {Vasiliev}\ \emph {et~al.}(2018)\citenamefont
  {Vasiliev}, \citenamefont {Volkova}, \citenamefont {Zvereva},\ and\
  \citenamefont {Markina}}]{Vasiliev:2018:}%
  \BibitemOpen
  \bibfield  {author} {\bibinfo {author} {\bibfnamefont {A.~N.}\ \bibnamefont
  {Vasiliev}}, \bibinfo {author} {\bibfnamefont {O.~S.}\ \bibnamefont
  {Volkova}}, \bibinfo {author} {\bibfnamefont {E.~A.}\ \bibnamefont
  {Zvereva}},\ and\ \bibinfo {author} {\bibfnamefont {M.~M.}\ \bibnamefont
  {Markina}},\ }\href@noop {} {\emph {\bibinfo {title} {Low dimensional
  magnetism}}}\ (\bibinfo  {publisher} {Fizmatlit},\ \bibinfo {address}
  {Moscow},\ \bibinfo {year} {2018})\BibitemShut {NoStop}%
\bibitem [{\citenamefont {Baxter}(1982)}]{Baxter:1982}%
  \BibitemOpen
  \bibfield  {author} {\bibinfo {author} {\bibfnamefont {R.~J.}\ \bibnamefont
  {Baxter}},\ }\href@noop {} {\emph {\bibinfo {title} {Exactly solved models in
  statistical mechanics}}}\ (\bibinfo  {publisher} {Academic Press},\ \bibinfo
  {address} {London},\ \bibinfo {year} {1982})\BibitemShut {NoStop}%
\bibitem [{\citenamefont {Kramers}\ and\ \citenamefont
  {Wannier}(1941)}]{Kramers:1941:1}%
  \BibitemOpen
  \bibfield  {author} {\bibinfo {author} {\bibfnamefont {H.~A.}\ \bibnamefont
  {Kramers}}\ and\ \bibinfo {author} {\bibfnamefont {G.~H.}\ \bibnamefont
  {Wannier}},\ }\href {https://doi.org/10.1103/PhysRev.60.252} {\bibfield
  {journal} {\bibinfo  {journal} {Phys. Rev.}\ }\textbf {\bibinfo {volume}
  {60}},\ \bibinfo {pages} {252} (\bibinfo {year} {1941})}\BibitemShut
  {NoStop}%
\bibitem [{\citenamefont {Domb}(1960)}]{Domb:1960}%
  \BibitemOpen
  \bibfield  {author} {\bibinfo {author} {\bibfnamefont {C.}~\bibnamefont
  {Domb}},\ }\href {https://doi.org/10.1080/00018736000101189} {\bibfield
  {journal} {\bibinfo  {journal} {Adv. Phys.}\ }\textbf {\bibinfo {volume}
  {9}},\ \bibinfo {pages} {149} (\bibinfo {year} {1960})}\BibitemShut {NoStop}%
\bibitem [{\citenamefont {Katsura}\ and\ \citenamefont
  {Ohminami}(1972)}]{Katsura:1972}%
  \BibitemOpen
  \bibfield  {author} {\bibinfo {author} {\bibfnamefont {S.}~\bibnamefont
  {Katsura}}\ and\ \bibinfo {author} {\bibfnamefont {M.}~\bibnamefont
  {Ohminami}},\ }\href {https://doi.org/10.1088/0305-4470/5/1/014} {\bibfield
  {journal} {\bibinfo  {journal} {J. Phys. A: Gen. Phys.}\ }\textbf {\bibinfo
  {volume} {5}},\ \bibinfo {pages} {95} (\bibinfo {year} {1972})}\BibitemShut
  {NoStop}%
\bibitem [{\citenamefont {Pokrovskii}\ and\ \citenamefont
  {Uimin}(1982)}]{Pokrovskii:1982:}%
  \BibitemOpen
  \bibfield  {author} {\bibinfo {author} {\bibfnamefont {V.~L.}\ \bibnamefont
  {Pokrovskii}}\ and\ \bibinfo {author} {\bibfnamefont {G.~V.}\ \bibnamefont
  {Uimin}},\ }\href@noop {} {\bibfield  {journal} {\bibinfo  {journal} {Sov.
  Phys. JETP}\ }\textbf {\bibinfo {volume} {55}},\ \bibinfo {pages} {950}
  (\bibinfo {year} {1982})}\BibitemShut {NoStop}%
\bibitem [{\citenamefont {Price}\ and\ \citenamefont
  {Yeomans}(1984)}]{Price:1984}%
  \BibitemOpen
  \bibfield  {author} {\bibinfo {author} {\bibfnamefont {G.~D.}\ \bibnamefont
  {Price}}\ and\ \bibinfo {author} {\bibfnamefont {J.}~\bibnamefont
  {Yeomans}},\ }\href {https://doi.org/10.1107/S0108768184002469} {\bibfield
  {journal} {\bibinfo  {journal} {Acta Cryst.~B}\ }\textbf {\bibinfo {volume}
  {40}},\ \bibinfo {pages} {448} (\bibinfo {year} {1984})}\BibitemShut
  {NoStop}%
\bibitem [{\citenamefont {Zarubin}\ \emph {et~al.}(2019)\citenamefont
  {Zarubin}, \citenamefont {Kassan-Ogly}, \citenamefont {Proshkin},\ and\
  \citenamefont {Shestakov}}]{Zarubin:2019:}%
  \BibitemOpen
  \bibfield  {author} {\bibinfo {author} {\bibfnamefont {A.~V.}\ \bibnamefont
  {Zarubin}}, \bibinfo {author} {\bibfnamefont {F.~A.}\ \bibnamefont
  {Kassan-Ogly}}, \bibinfo {author} {\bibfnamefont {A.~I.}\ \bibnamefont
  {Proshkin}},\ and\ \bibinfo {author} {\bibfnamefont {A.~E.}\ \bibnamefont
  {Shestakov}},\ }\href {https://doi.org/10.1134/S106377611904006X} {\bibfield
  {journal} {\bibinfo  {journal} {J. Exp. Theor. Phys.}\ }\textbf {\bibinfo
  {volume} {128}},\ \bibinfo {pages} {778} (\bibinfo {year}
  {2019})}\BibitemShut {NoStop}%
\bibitem [{\citenamefont {Price}(1983)}]{Price:1983}%
  \BibitemOpen
  \bibfield  {author} {\bibinfo {author} {\bibfnamefont {G.~D.}\ \bibnamefont
  {Price}},\ }\href {https://doi.org/10.1007/BF00309588} {\bibfield  {journal}
  {\bibinfo  {journal} {Phys. Chem. Miner.}\ }\textbf {\bibinfo {volume}
  {10}},\ \bibinfo {pages} {77} (\bibinfo {year} {1983})}\BibitemShut {NoStop}%
\bibitem [{\citenamefont {Barreto}\ and\ \citenamefont
  {Yeomans}(1985)}]{Barreto:1985}%
  \BibitemOpen
  \bibfield  {author} {\bibinfo {author} {\bibfnamefont {M.}~\bibnamefont
  {Barreto}}\ and\ \bibinfo {author} {\bibfnamefont {J.}~\bibnamefont
  {Yeomans}},\ }\href {https://doi.org/10.1016/0378-4371(85)90157-8} {\bibfield
   {journal} {\bibinfo  {journal} {Physica A}\ }\textbf {\bibinfo {volume}
  {134}},\ \bibinfo {pages} {84} (\bibinfo {year} {1985})}\BibitemShut
  {NoStop}%
\bibitem [{\citenamefont {Selke}\ \emph {et~al.}(1985)\citenamefont {Selke},
  \citenamefont {Barreto},\ and\ \citenamefont {Yeomans}}]{Selke:1985}%
  \BibitemOpen
  \bibfield  {author} {\bibinfo {author} {\bibfnamefont {W.}~\bibnamefont
  {Selke}}, \bibinfo {author} {\bibfnamefont {M.}~\bibnamefont {Barreto}},\
  and\ \bibinfo {author} {\bibfnamefont {J.}~\bibnamefont {Yeomans}},\ }\href
  {https://doi.org/10.1088/0022-3719/18/14/007} {\bibfield  {journal} {\bibinfo
   {journal} {J. Phys. C: Solid State Phys.}\ }\textbf {\bibinfo {volume}
  {18}},\ \bibinfo {pages} {L393} (\bibinfo {year} {1985})}\BibitemShut
  {NoStop}%
\bibitem [{\citenamefont {Fisher}\ and\ \citenamefont
  {Selke}(1981)}]{Fisher:1981}%
  \BibitemOpen
  \bibfield  {author} {\bibinfo {author} {\bibfnamefont {M.~E.}\ \bibnamefont
  {Fisher}}\ and\ \bibinfo {author} {\bibfnamefont {W.}~\bibnamefont {Selke}},\
  }\href {https://doi.org/10.1098/rsta.1981.0156} {\bibfield  {journal}
  {\bibinfo  {journal} {Phil. Trans. R. Soc. A}\ }\textbf {\bibinfo {volume}
  {302}},\ \bibinfo {pages} {1} (\bibinfo {year} {1981})}\BibitemShut {NoStop}%
\bibitem [{\citenamefont {Zarubin}\ \emph {et~al.}(2016)\citenamefont
  {Zarubin}, \citenamefont {Kassan-Ogly},\ and\ \citenamefont
  {Proshkin}}]{Zarubin:2016}%
  \BibitemOpen
  \bibfield  {author} {\bibinfo {author} {\bibfnamefont {A.~V.}\ \bibnamefont
  {Zarubin}}, \bibinfo {author} {\bibfnamefont {F.~A.}\ \bibnamefont
  {Kassan-Ogly}},\ and\ \bibinfo {author} {\bibfnamefont {A.~I.}\ \bibnamefont
  {Proshkin}},\ }\href {https://doi.org/10.4028/www.scientific.net/MSF.845.122}
  {\bibfield  {journal} {\bibinfo  {journal} {Mater. Sci. Forum}\ }\textbf
  {\bibinfo {volume} {845}},\ \bibinfo {pages} {122} (\bibinfo {year}
  {2016})}\BibitemShut {NoStop}%
\end{thebibliography}%

\end{document}